\documentclass[amsmath,amssymb]{revtex4}
\usepackage[dvips]{graphicx}
\usepackage{epsfig}
\usepackage{subfigure}
\newcommand{\ket}[1]{\left| #1 \right\rangle}

\newcommand{\<}{\langle}
\renewcommand{\>}{\rangle}
\newcommand{\be}{\begin{equation}}
\newcommand{\ee}{\end{equation}}
\newcommand{\bea}{\begin{eqnarray}}
\newcommand{\eea}{\end{eqnarray}}

\graphicspath{%
    {converted_graphics/}
    {/}
}
\begin{document}
\title{Non-Hermitian exciton dynamics in  a  photosynthetic unit system}
\author{ A. Thilagam} 
\email{thilaphys@gmail.com}
\address{Information Technology, Engineering and Environment, 
Mawson Institute,
University of South Australia, Australia
 5095} 

\begin{abstract}
The non-Hermitian quantum dynamics of excitonic energy  transfer in
photosynthetic systems is investigated using a dissipative two-level dimer model.
The approach is based on the  Green's function formalism
which permits consideration of  decoherence and intersite transfer
 processes on comparable terms. The results indicate a combination 
of coherent and incoherent behavior at higher temperatures with the possibility of
  exceptional points occurring at the coherent-incoherent crossover regime at 
critical temperatures. When each dimer site is coupled  {\it equally}  to the environmental sources of 
dissipation, the excitonic wavepacket evolves with time with  a 
 coherent component,  which can be attributed 
to the indistinguishability of the sources of dissipation.
 The  time evolution characteristics of the
 B850 Bchls dimer system is analysed using  typical parameter
 estimates  in photosynthetic systems, and the
 quantum brachistochrone passage times are obtained for a range of parameters.

\noindent PACS-number(s): 71.35.-y, 03.65.Yz, 03.67.Mn
\end{abstract}
\maketitle

\section{Introduction}

The theory of excitonic energy  transfer  has been a topic of
 interest  over several decades  in  light harvesting systems 
(LHS) \cite{p0,fors,p1,p1a,p2,p3,may,macro,lloyd,silbey,ghosh,olbri,pala1,pala2,pala}.
The Fenna$-$Matthews$-$Olson (FMO) complex 
of the green sulfur bacteria \cite{flem,fenn,engel,renrat} 
 constitutes the prototypical LHS  for modelling photosynthetic activities \cite{reng,brug,lloyd},
mainly  due to its generic features of energy transfer also seen in other photosynthetic systems \cite{p2,p3,may}.
In the FMO complex,  two types of light-harvesting molecular systems complexes, known as LH1
and LH2,  perform different roles. The LH1 is directly linked with  the
reaction center (RC)  unlike the LH2 complex which interacts with the
RC  via  LH1. The main attraction in light harvesting systems is the 
exceptionally high efficiencies at which
 excitation  propagates between the  light harvesting complexes before reaching the  reaction center (RC) pigment–
protein complex \cite{p1,p3,may}. Infact, the quantum efficiency in transfer at low illumination intensities 
reaches close to a value of unity.
  The  importance of mimicking the transfer processes seen in
LHS has obvious applications in  artificial light harvesting system such as Ruthenium based  
complexes \cite{ener1a,ener1b,ener1c}. To this end, accurate knowledge of
the underlying mechanisms of energy transport and the process by which 
 excess energy in  excited  states is transferred  to their surroundings in 
 LHS   is critical  to the development of
 potential sources of energy generation \cite{ener1,ener2,ener3,ener4}.

Early experimental work in 1960 by   Chance and Nishimura \cite{p1a} 
 showed that  some level of photosynthetic activity 
occurred in  chlorophyll systems close to room temperatures (300 K).
The occurrence of quantum coherence during photosynthesis 
was not obvious due to the level of precision of  the  apparatus employed at that time. 
However recent  progress in experimental techniques such as
two-dimensional Fourier transform electronic spectroscopy (2DFTES)
\cite{flem,engel,collini,flemB}  has confirmed  that  quantum coherence is indeed conserved for 
relatively long times (up to 1 picosecond) in excited states.
Using 2DFTES \cite{engel}, the FMO complex of green sulfur bacteria
showed  a surprisingly long coherence time  of about 700 fs at 77 K,
and also at room temperatures (277 K) in a related work \cite{pani}.
These ambient temperatures  are generally
 considered adverse for  quantum features such as superpositions and non-classical correlations 
to be maintained for a reasonable period of time. 
However the experimental results \cite{flem,engel,collini,pani,flemB} appear to suggest the critical role 
played by  non-local quantum effects in energy transfer mechanisms 
in LHS.  To this end, the application of advanced measurement tools
involving spectrally resolved, 4-wave mixing measurements \cite{segale} 
is expected to introduce greater depth to the study
of quantum correlations in   photosynthetic systems which undergo both  
decoherence and dissipation due to  interactions  with the environment.

Galve et. al. \cite{galve} showed recently that for a system of two interacting 
and parametrically driven harmonic oscillators undergoing dissipation, 
entanglement can exists at any temperatures. Although this specific quantum 
system appear distinct from quantum models which describe LHS, the 
general consensus  is that the slower processing speed inherent in classical models
is inadequate for LHS compared to the  processing attributes of quantum systems.
This viewpoint has been the focus of  many recent works
 \cite{fas,car,lloyd,sar,scholak,nazir,fass,nal} involving LHS. A variety of methods involving
the Redfield approach employing the   Born-Markov and secular 
approximations \cite{red,weiss,breu},  the
non-Markovian approach which involves solving the 
 integrodifferential equation  using  perturbation theory \cite{pot}, and  recently 
 the more sophisticated and numerically intensive 
reduced hierarchy equation approach \cite{hierac} have been employed to examine
the propagation of excitation in LHS. 
In   light-harvesting complexes with strong system-bath coupling, 
 a generalized Bloch–Redfield (GBR) equation approach
  was used   to show that optimal efficiency in 
energy transfer cannot be optimized with respect to the
 temperature and spatial–temporal correlations in noise \cite{caoIop}. 
Thus classical approaches to 
the excitonic transport which involve the  hopping model  \cite{p0,p2,fors}  have given way to 
  more sophisticated quantum mechanical models of exciton propagation \cite{lloyd,ghosh,hierac,caoIop,shaun}
 which incorporate  quantum coherence aspects during  the energy transfer process.

Green's functions are known to provide an  effective description  of the
quantum evolution of   non equilibrium quantum processes \cite{econ,mal,suna,Dav,thila}, and here
we use this approach  to examine the influence of  dissipative processes 
on the dynamical evolution of the exciton population in LHS.
While several works have considered
coherence and bath correlations \cite{johan,car,fas} 
as possible  factors for the high efficiencies of LHS, here we consider
two  features which may account for the quantum properties in LHS: 
 (a) the persistence of coherence  when the dissipative  coupling of 
each monomer  to its surrounding environment becomes equivalent to that 
of the adjacent monomer, and 
(b) the appearance of exceptional points at critical temperatures.
The simple model of the dimer is used to show that in the 
event that the coupling of each subsystem to the phonon bath or other sources of 
dissipation is equal, the overall time
evolution still retains a  coherent component, an idea that was first proposed by
Stafford and Barrett \cite{statt} in the context of the decay of 
super deformed nuclei systems.  This key feature  implies that a degree of  coherence  
is maintained even at temperatures that would otherwise be considered adverse for
 any superpositions of the quantum states. Accordingly we 
consider the maintenance of coherence, as quantified by the time duration 
of oscillations in the population difference at the two  site model,
 as a measure of the photosynthetic efficiency. In general,
the photon is absorbed by a network system of pigments and transported to the
reaction center (RC)  trap  so that efficiency is based on the success of being trapped at the RC. 
The connection between quantum coherence and photosynthetic efficiency is justified
as the loss of coherence disrupts the likelihood of the propagating exciton 
of reaching its destination due to decoherence processes.
 
Unlike earlier works on LHS, we  take advantage of the  non-Hermitian features inherent
in any open quantum system  which encounter dissipative 
forces. We also aim to examine the underlying factors which are linked  with the non-Hermitian
features and which may account for the observed  long-lived coherence in LHS. 
Open quantum systems with non-Hermitian components
 evolve in ways which are vastly  different 
 from quantum systems associated with a  purely Hermitian Hamiltonian.
For instance, the states associated with an Hermitian Hamiltonian  have 
long lifetimes, while those of the  non-Hermitian Hamiltonian
have a finite lifetime. Non-Hermitian resonances are associated with 
complex eigenvalues with  real (imaginary)  component
that yield  the energy (resonance width), and possess non-orthogonal  eigenvectors.
Thus states which are orthogonal  under the ordinary 
inner product in the Hermitian quantum space  are allowed to be non-orthogonal
in the  non-Hermitian case. The real  state energies  of 
a  Hermitian Hamiltonian give rise to the avoided level crossing \cite{tell} when
a single parameter is varied, which is not necessarily true in  the case 
of a non-Hermitian Hamiltonian.   The presence of non-Hermitian terms is critical to the occurrence of 
dynamical phase transitions as well. Moreover the appearance of degeneracies such as
 exceptional points (EPs)  is a unique feature of non-Hermitian quantum systems, 
and may assist in distinguishing classical and quantum modes of transport in LHS.

A topological defect like the exceptional point occurs when  two eigenvalues of an operator coalesce
as a result of change in selected system parameters, and  the two mutually orthogonal
 states merge into one self-orthogonal state, 
resulting in a singularity in the spectrum \cite{heiss}. 
The critical parameter values  at which the singularity appear  are called exceptional
points. A notable feature associated with exceptional points is the violation of the adiabatic theorem,
that is the switching of an initial eigenstate to another eigenstate when the
system parameters are altered adiabatically around the exceptional point.
In contrast, Hermitian systems exhibit only berry phases \cite{thilaberry}, with no change in the
initial eigenstate as the system parameters are altered adiabatically. Thus no
exceptional points can exist  in Hermitian systems and at degenerate points
 the eigenstates exist only  in a two-dimensional  subspace  spanned
by vectors. EPs have been observed in experiments
 involving microwave billiards \cite{microexpt,b1,b2},
semiconductors cavities \cite{cav} and  photo-dissociated 
 vibronic resonance states of the H$_2^{+}$ molecular ion \cite{lefe} .
Exceptional points may be associated with   the transition or crossover points at which
coherent to incoherent tunnelling occur in open quantum systems. The detection of the
exceptional  points are still under active investigation \cite{lefe2}. One possible technique involves
 the detection of the switching process, in which two states around an exceptional point
swap states and thus behave distinctly from other states which remain unchanged
 when the selected  parameter is changed adiabatically \cite{carta}. However 
the experimental detection of such topological changes still remains a challenge.

This work is organized as follows: In Sec. (\ref{time}) we present the
theory of the exciton transfer using Green's function formalism and analyse the
influence of  the intersite coupling energy and dissipation rates on the 
coherence properties of the  exciton in a simple dimer system.
In Sec. (\ref{excep}), we examine the photosynthetic qubit system
and discuss the conditions under which exceptional points appear.
In Sec. (\ref{bch}), the dimer model is applied to the  B850 Bchls system
and numerical  estimates of the time scales of coherent  oscillations in the  exciton 
population difference are obtained. Estimates of the critical temperatures 
at which exceptional points occur are also evaluated for a range of environmental 
dissipation strength differences. Lastly, in Sec. (\ref{brac}),
the quantum brachistochrone passage times are obtained for a range
of parameters   in photosynthetic systems. 
A brief discussion and conclusion are also provided
in  Sec. (\ref{brac}).

\section{\label{time} Exciton transfer using Green's function formalism}
The Hamiltonian describing the propagation of 
the exciton in molecular systems is 
based on the tight-binding model \cite{Dav}
\be
\label{exH}
\hat{H}_{ex}= \sum_{ l}\left[ {\Delta E} +
{\sum_{ m \neq l}} D_{ l,m} \right ] \; B_{ l}^\dagger B_{ l}+
\sum_{ m \neq l} V B_{ l}^\dagger B_{ m}
\ee
where  $B_j^\dagger$ ($B_j$) is the 
 creator (annihilation) exciton operator  at   site $j$.
 $\Delta E$, the on-site excitation energy and  
$D_{ l,m}$ is the dispersive interaction matrix element
which determines  the   energy difference between a pair of 
excited electron and hole at a molecular site and ground state electrons
at neighboring sites \cite{Dav}.  $V$  the 
exciton transfer matrix element between molecular sites at $l$
and $m$. In  the case of the dimer system with just two coupled  sites
 (labeled $l$ and $m$),  Eq. (\ref{exH}) is greatly simplified 
\be
\label{exHsim}
\hat{H}_{ex}= E_l^0  B_{ l}^\dagger B_{ l}+ E_m^0 B_{ m}^\dagger B_{ m}+
V B_{ l}^\dagger B_{ m}
\ee
where $E_l, E_m$ are the exciton energies at sites $l,m$ and 
 the subscript $0$ denotes the absence of lattice site fluctuations.

In the presence of a non-Hermitian decay terms and 
 phonon bath reservoirs at each site, the
  exciton dynamics of the  dimer system is determined by the 
following  Hamiltonian 
\bea
\label{exHim}
\hat{H}_{T}&=& \hat{H}_{ex} + \hat{H}_d+ \hat{H}_{p}
+\hat{H}_{ep} \\
 \label{dis}
\hat{H}_{d}&=& i \sum_{ j=l,m} \zeta_j B_{j}^\dagger B_{j} \\
\label{phon}
\hat{H}_{p}&=& \sum_{ j=l,m}\sum_{ q} \hbar \omega({ q,j})b_{ q,j}^\dagger b_{ q,j}
\\
\label{exphon}
\hat{H}_{ep}&=&  N^{-1/2} \sum_{ j=l,m} \sum_{ q} \;F_j(q) B_{j}^\dagger
B_{j} (b_{- q,j}^\dagger +b_{ q,j}) 
 \eea
where the isolated exciton Hamiltonian, $\hat{H}_{ex}$ is given in  Eq. (\ref{exHsim}).
We allow the   intersite   tunnelling  amplitude $V$ to be influenced by
a  phonon correlated environment via  a Franck-Cordon (FC) factor  \cite{legg} 
\be
\label{fc}
V_r= V \; \exp \left[-\int_0^\infty \frac{J(\omega)}{\omega^2}
 \; \coth\left(\frac{\hbar \omega}{2 k_B T}\right) {\rm d} \omega \right ]
\ee
 where the spectral density function is  given by $J(\omega)$=
=$\sum_{ q} |F_a(q)|^2 \; \delta(\omega-\omega_q)$. We assume that the correlated
exciton-phonon interaction term $F_a(q)$ to be the average ($ \frac{1}{2}[F_l(q)+F_m(q]$) of 
the interactions  at sites, $l,m$.
The argument of the exponential term in Eq. (\ref{fc}) is 
known as the FC  factor \cite{legg} and its dependency on the temperature
varies according to the form chosen for $J(\omega)$.

The non-Hermitian dissipative Hamiltonian in  Eq. (\ref{dis}),
$\hat{H}_{d}$ is associated with  $\zeta_j$,  the decay rate of the exciton at each site $j=l,m$.
For simplicity in numerical analysis, we have incorporated  several processes
such as exciton annihilation due to  recombination and trapping effects
into $\zeta_j$, with the requirement that the decay rate is dependent on the site $j$.
Due to differences in  environmental conditions at the two sites of the dimer $\zeta_l \neq \zeta_m$, however 
the possibility that  $\zeta_l \approx \zeta_m$  cannot be excluded.
$\hat{H}_{p}$ denotes the sum of
phonon energies at the two sites and $b_{ q,j}^\dagger (b_{ q,j})$ is the creation (annihilation) phonon
operator associated with the phonon bath at site $j$, with
 frequency $\omega({ q})$ and  wavevector ${ q}$. $\hat{H}_{ep}$ represents the sum of 
the exciton-phonon interactions at the two sites, and  $F_j(q) $ quantifies the interaction
strength which is assumed to be linear in the phonon operators and which is dependent only on the
phonon wavevector $q$. We consider that both
dissipation and dephasing  processes  occur as a consequence of 
exciton-phonon interactions. The 
 dissipative mechanisms associated with the exciton-phonon 
interactions   introduce a non-Hermitian term in the 
Green's function (see Eq. (\ref{herm2}) below) which account for the
irreversible loss in the exciton population.
 While the  dephasing term leaves the overall exciton population unchanged,
it  contributes to incoherence by changing the non-diagonal
components  of the density matrix.

The Green's function for an exciton at time $t$ which yields
the dynamical measure of the amplitude of  the electron (hole)
to propagate forward (backward) in time is given by \cite{Dav}
\be
\label{g1}
G_{l,m}(t)=-i  \Theta(t) \<\{B_l(t) B_{m}^\dagger\}|\>
\ee
where $\Theta(t)$ denotes the step function and the exciton Green's function
is averaged over the quantum mechanical properties of the propagating excitonic
wavefunction that is dependent on the phonon bath and contribution from the dissipative terms.
Following the approach by  Stafford and Barrett \cite{statt},
the Fourier transform  $G_{l,m}(E)$ is given in terms of the energy $E$ 
as  $G_{l,m}(E)=\int_{-\infty}^{\infty} dt\, G_{l,m}(t)\, e^{iEt}$.
For  the simple case of the dimer system with two coupled  sites of  energies $E_l,E_m$, 
the inverse of the   Green's function $G_{l,m}(E )$ is obtained as
\be
\label{herm1}
G_{l,m}^{-1}(E)= \left[ \begin{array}{cc} E-E_l-\Delta_l+i \eta+ & -V_r \\ -V_r & 
E-E_m-\Delta_m+i \eta+ \end{array} \right]- \frac{1}{2}\Sigma.
\ee
where $\eta$ is a very  small number and $\Delta_j$ is the shift in the exciton
energy due to its interaction with phonons. This polaronic energy shift  can be evaluated using
 standard techniques based on the polaron model \cite{mah}. Without loss in generality, the polaronic 
shift term can be incorporated  as a constituent of the site energy $E_l$ or $E_m$, and therefore
we drop this term from now on. 
 $\Sigma$ is the self-energy due to dissipative interactions with the phonon bath
and other processes associated with  exciton recombination, annihilation and trapping
which is expressed as
\be
\label{herm2}
\Sigma =  \left( \begin{array}{cc} - i (\gamma_{ph,l}+ \gamma_{d,l})& 0 \\ 0 & 
-i(\gamma_{ph,m}+ \gamma_{d,m}) \end{array} \right).
\ee
Here we distinguish the phonon-related dissipative term $\gamma_{ph,j}$  
from the dissipative term arising from non-phonon related 
recombination and trapping effects which is denoted by $\gamma_{d,j}$. 
$\gamma_{ph,j}$  at site $j$ is obtained as
\be
\gamma_{ph,j}=N^{-1} \sum_{ q} |F_j(q)|^2 \delta(\hbar \omega +E_m-E-\Delta p)
 \ee
where $E_m$ is the mean energy of the exciton band, $E$ is the exciton energy
and $\Delta p$ can be interpreted as the dissipative  or trap  depth \cite{Thilaprb} which 
results in the irreversible loss of the exciton. Both $E_m$ and $\Delta_p$ appear as phenomenological
constants  and their associated values are unavailable for the  FMO complex. For the purpose
of obtaining numerical values in the next section,  we  select a range of  values for the cumulative
term ($\gamma_{ph,j}$+$\gamma_{d,j}$)  and examine its   
 influence on the coherence properties of the excitonic dimer.


Using the inversion procedure adopted by Stafford and Barrett \cite{statt},
we obtain 
\bea
 \label{green}
G(E)&=& \left(\left[E-E_l+\frac{i}{2}\bar{\gamma}_l\right]\;
\left[E-E_m+\frac{i}{2}\bar{\gamma}_m \right]-{V_r}^2\right)^{-1} 
\\ \nonumber
\times &&
\left[\begin{array}{cc} E-E_m+\frac{i}{2}\bar{\gamma}_m& V_r \\ V_r & 
E-E_l+\frac{i}{2}\bar{\gamma}_l\end{array} \right].
\eea
where ${\gamma}_m= \gamma_{ph,m}+ \gamma_{d,m}$
and ${\gamma}_l= \gamma_{ph,l}+ \gamma_{d,l}$.
We consider that the exciton in the simple dimer system
 is initially localized at site $j=l$ at time $t=0$, thus the
probability $P_{ll}$ that the exciton  remains at site $j=l$ is determined using
$P_{ll}(t)=|G_{11}(t)|^2$.  Explicit expressions for $P_{ll}(t)$ and the probability that
the exciton has propagated to site $j=m$, $P_{lm}(t)$
can be obtained for equal site energies, $E_l=E_m$  in the 
 coherent tunneling regime ($2 V_r > \gamma^{\star}$) 
\begin{eqnarray}
\nonumber
P_{ll} &=&  e^{-\bar{\gamma} t}
\left[\cos\frac{\Omega}{2} t- \frac{\gamma^{\star}}{\Omega}\sin \frac{\Omega}{2} t \right]^2 
\\ \label{eq:co}
 P_{lm} &=&  e^{-\bar{\gamma} t} \frac{4 {V_r}^2}{\Omega^2}  \sin^2 \frac{\Omega}{2} t,
\end{eqnarray}
where  $\Omega =(4 {V_r}^2 -{\gamma^{\star}}^2)^{1/2}$,
$\gamma^{\star}= \frac{1}{2}(\gamma_m - \gamma_l)$ and 
$\bar{\gamma}= \frac{1}{2}(\gamma_m + \gamma_l)$.
While the contribution of the dissipative term, $\bar{\gamma}$  appears  in the  exponential
function (see Eq.(\ref{eq:co})), the dimer can be seen to still exhibit coherence with  an effective
 Rabi oscillation that is  influenced by the environment.
The system however undergoes incoherent tunneling with loss of 
Rabi oscillations at  $2 V_r < \gamma^{\star}$ and 
we obtain
\begin{eqnarray}
\nonumber
P_{ll} &=&  e^{-\bar{\gamma} t}
\left[\cosh \frac{\Omega}{2} t- \frac{\gamma^{\star}}{\Omega}\sinh \frac{\Omega}{2} t \right]^2 
\\ \label{eq:inco}
 P_{lm} &=&  e^{-\bar{\gamma} t} \frac{4 {V_r}^2}{\Omega^2}  \sinh^2 \frac{\Omega}{2} t,
\end{eqnarray}
Due to the presence of an non-Hermitian term in Eqs.(\ref{herm1}), (\ref{herm2}),
 the total probabilities, $P_{ll}$+$P_{lm} \le 1$ is not conserved and there is 
  loss of normalization which is  dependent on the dissipative terms, ${\gamma}_l$
 and ${\gamma}_m$.

We note that in the case of unequal energies, $E_l \neq E_m$,  the dimer
system exhibit both incoherent and coherent characteristics with a combination
of the terms which appear in Eqs.(\ref{eq:co}) and (\ref{eq:inco}). In the 
event of equivalent  dissipative couplings at each subsystem,
 we   obtain ${\gamma}_l={\gamma}_m$, $\gamma^{\star}$=0,
and the imaginary term present  in  $\Omega$ vanishes as noted earlier in Ref.\cite{carda} 
for a study involving quantum dot systems.
Despite the presence of non-zero
 $\gamma_l$ and $\gamma_m$,  Rabi oscillation of
 frequency $\Omega=\sqrt{4 {V_r}^2+(E_l-E_m)^2}$ occurs.
The subsystems of the dimer can be considered to be unmeasured by the
environmental sources when there is 
equivalent couplings to the dissipation channels, an observation 
that was first noted by Stafford and Barrett \cite{statt} for super deformed nuclei systems.
In this regard the indistinguishability of the source of decoherence preserves  the coherence of the excitonic dimer 
in photosynthetic systems.

\begin{figure}[htp]
  \begin{center}
\subfigure{\label{aa}\includegraphics[width=4cm]{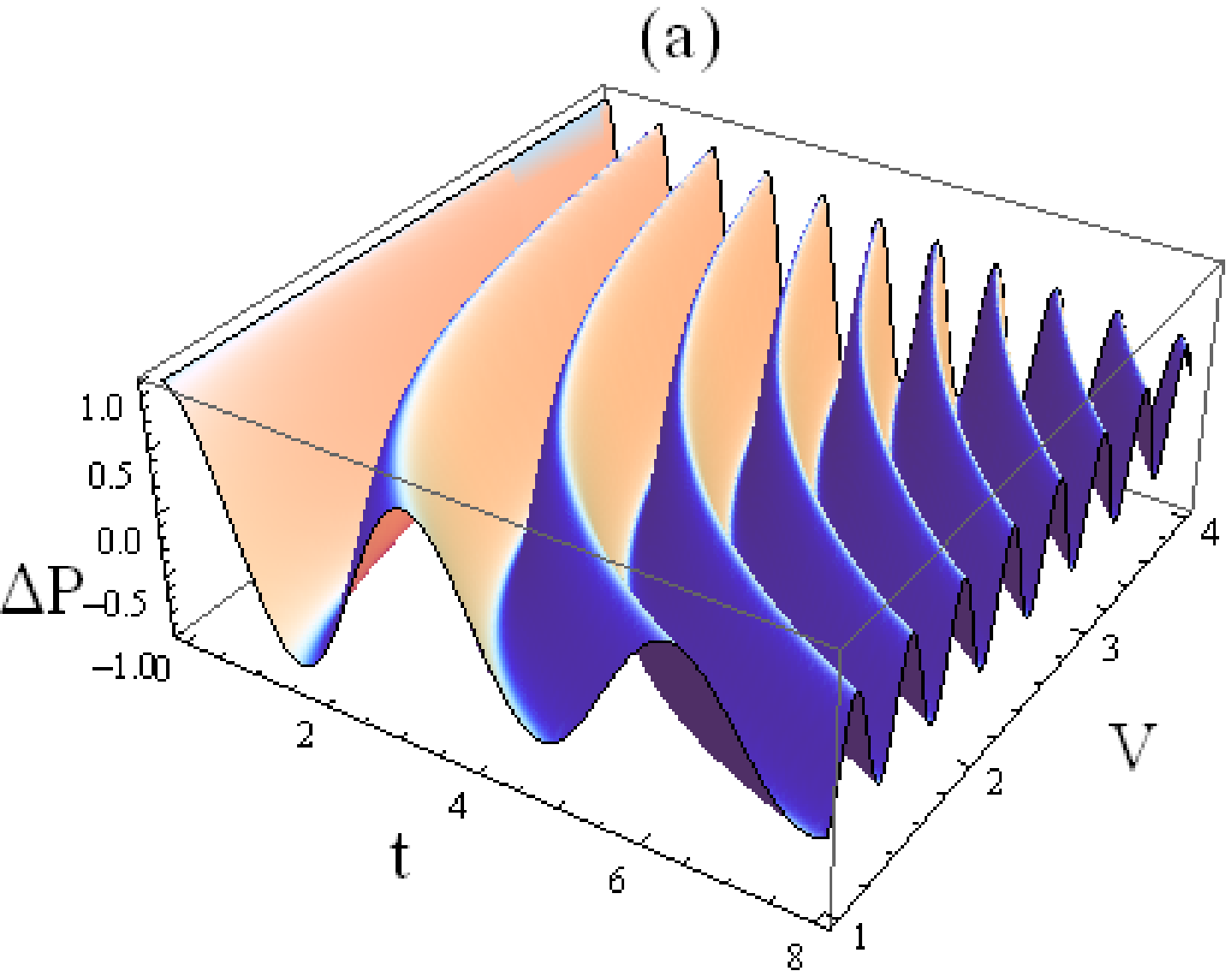}}\vspace{-1.1mm} \hspace{1.1mm}
\subfigure{\label{bb}\includegraphics[width=4cm]{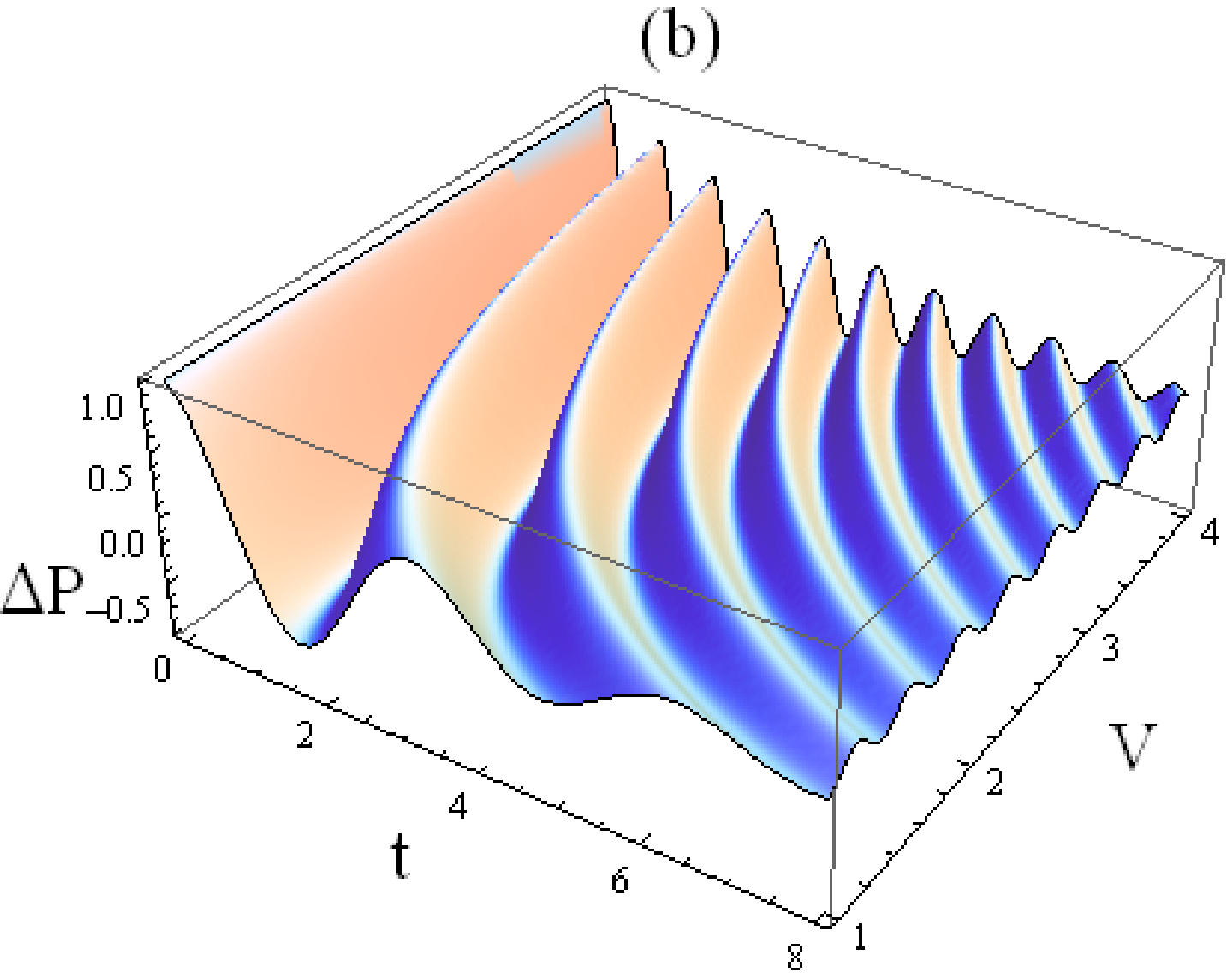}}\vspace{-1.1mm} \hspace{1.1mm}
\subfigure{\label{bb}\includegraphics[width=4cm]{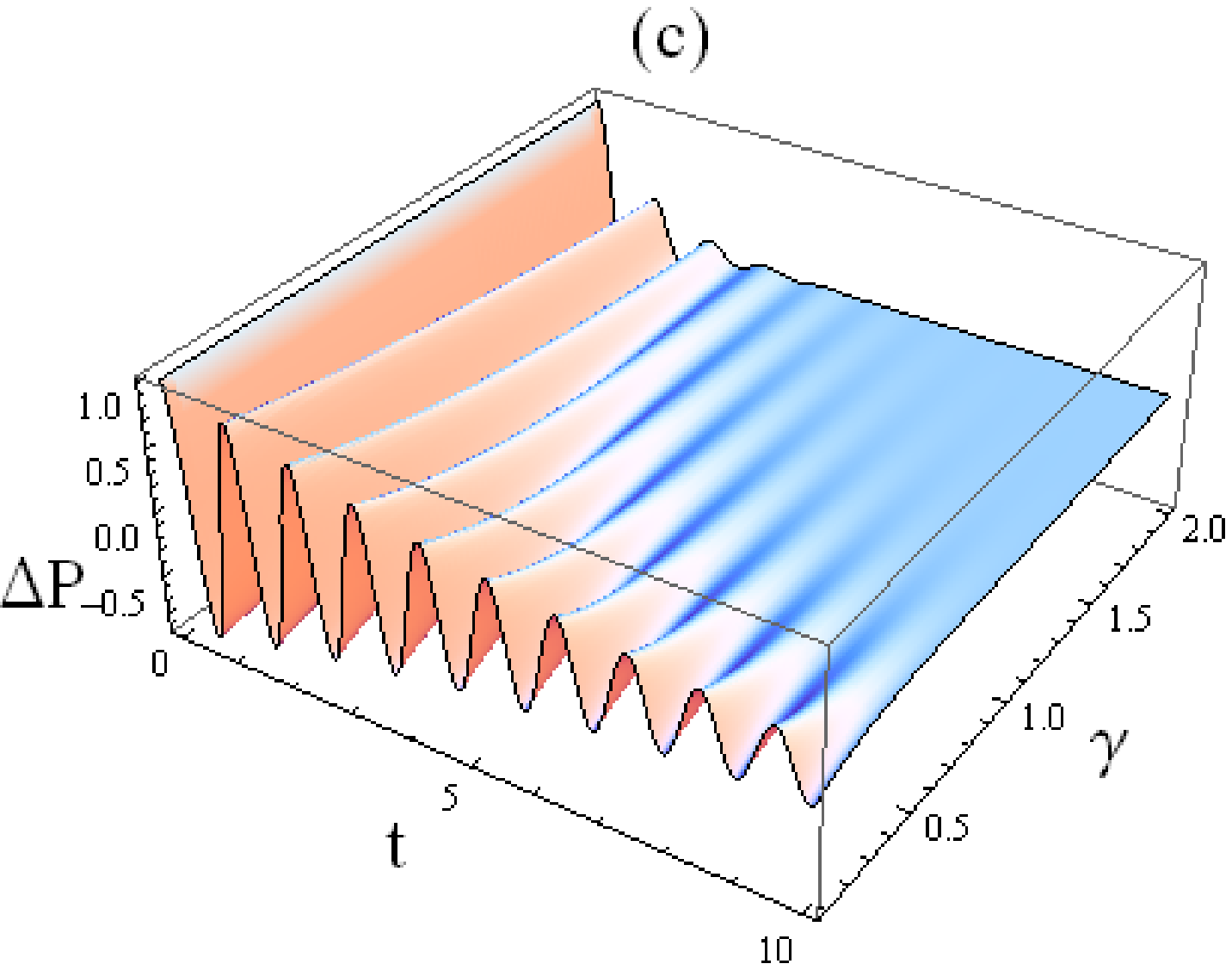}}\vspace{-1.1mm} \hspace{1.1mm}
     \end{center}
  \caption{(a) The exciton population difference, $\Delta P$=$P_{ll}-P_{lm}$ 
as a function of time $t$,  and the  bare intersite coupling energy, $V$ at dissipation rates 
${\gamma}_m$=${\gamma}_l$=0.1 for the degenerate case ($E_l$ = $E_m$). 
The units are chosen such that $\hbar$=1,
and the phonon bath response time, $\omega_0$=1.  \\
(b)  $\Delta P$=$P_{ll}-P_{lm}$
as a function of time $t$, the  bare intersite coupling energy, $V$ at dissipation rates 
${\gamma}_m$=0.1, ${\gamma}_l$=0.5. \\
(c) $\Delta P$=$P_{ll}-P_{lm}$ as function of time $t$ and dissipation rate ${\gamma}_m$,
at $\gamma_l$=0.2 and $V$=3.
}
\label{tau}
\end{figure}

In order to examine  the effect of the intersite coupling energy, $V$
and the dissipation rates ${\gamma}_m$, $\gamma_l$ on the
exciton dynamics, we consider the bare intersite energy $V$ without  the 
Franck-Cordon (FC) factor  \cite{legg}. 
Figure~\ref{tau}a,b,c shows the  coherence properties as reflected in the 
exciton population difference, $\Delta P$=$P_{ll}-P_{lm}$ as a function of time 
and the bare intersite coupling energy, $V$.
 We note that an increase in the strength of intersite energy $V$ increases
the time period over which the population difference, $\Delta P$ which is a signature 
of coherence, is maintained. As illustrated in
Figure~\ref{tau}c,  the gradual increase in one of the dissipation
rate ${\gamma}_m$ erodes the coherence in the dimer system as time progresses.

\section{\label{excep} Photosynthetic qubits and appearance of exceptional points}
Following Eq.(\ref{green}) and the form of the occupation probabilities in Eqs.(\ref{eq:co})  and 
(\ref{eq:inco}), the  symmetric and anti-symmetric states of the dimer system at the resonance point ($E_l$=$E_m$) 
is obtained as
\bea
\nonumber
\ket{\chi_{\rm s}(t)}  & = &  e^{-\bar{\gamma} t/2} \left(\cos\frac{\Omega}{2} t-i \cos \theta \sin \frac{\Omega}{2} t \right)
 \ket{\bf l} \; \\ \nonumber&&+ i  e^{-\bar{\gamma} t/2} \sin \theta \sin\frac{\Omega}{2}  t \ket{\bf m} \\ \nonumber
\ket{\chi_{\rm a}(t)}  & = &  e^{-\bar{\gamma} t/2}  \left(\cos\frac{\Omega}{2} t-i \cos \theta\sin \frac{\Omega}{2} t \right)\ket{\bf l} \;
 \\ \nonumber && -i  e^{-\bar{\gamma} t/2} \sin \theta \sin\frac{\Omega}{2} t \ket{\bf m},
\\
\label{Eigen}
\eea
where   $\cos \theta$=$\frac{i \gamma^{\star}}{\Omega}$ and
$\Omega$ and $\gamma^{\star}$ are defined below
 Eqs.(\ref{eq:co}).  The excitonic qubits states in Eqs.(\ref{Eigen}) are coded using  
the relative position of the exciton via the  
basis set, ($\ket{\bf l}, \ket{\bf m}$) 
\begin{eqnarray}
\nonumber
\label{qstates}
\ket{\bf l}  & = &  \ket{\bf \Xi_{\rm l}} \otimes 
 \ket{\bf 0}_m \\
\ket{\bf m}  & = & \ket{\bf 0}_l \otimes \ket{\bf \Xi_{\rm m}},
\end{eqnarray}
where  $\ket{\bf \Xi_{\rm l}}$ ($\ket{\bf \Xi_{\rm m}}$) denote the excitonic
state at site $l$ ($m$), and the state  $\ket{\bf 0}_l$ ($\ket{\bf 0}_m$)
 denote   the  ground states  at site $l$  ($m$) (the absence of exciton).
Other than the qubit states, $\ket{\chi_{\rm s}(t)}$ and $\ket{\chi_{\rm a}(t)}$, 
 entangled states of the form  $\ket{\bf 0}_l \ket{\bf 0}_l$ and  
$\ket{\bf \Xi_{\rm l}}  \ket{\bf \Xi_{\rm m}}$ may also result in the dimer system. 

At unequal energies, $E_l \neq E_m$, the states  equivalent to the
symmetric and asymmetric states in Eq.(\ref{Eigen}) possess    eigenenergies of the form
\be
\label{compE}
E_\pm={\bar{E}_l+
\bar{E}_m} \pm \sqrt{\left( {\bar{E}_m-
\bar{E}_l} \right)^2 +4 {V_r}^2}.
\ee
where the complex energy levels $\bar{E_l}$=$E_l-i {\gamma}_l$ 
and $\bar{E_m}$=$E_m-i {\gamma}_m$  and the dissipative terms 
${\gamma}_m$ and ${\gamma}_m$  are specified below  Eq.(\ref{eq:co}).
The excitonic qubit oscillates coherently between the
two dots with the complex Rabi frequency $E_{+} - E_{-}$=$\bar{\omega}$=
$2 \left [ {[({\bar{E}_m- \bar{E}_l})]^2 + 4 {V_r}^2} \right]^{1/2}$.
By setting $E_l =E_m$, the eigenenergies corresponding to Eq.(\ref{Eigen})
can be easily evaluated. The real component of $\bar{\omega}$
correspond to Rabi oscillations  while the imaginary
component correspond to the rate of incoherent tunneling  \cite{carda}.
The role of a complex Rabi  frequency has been discussed \cite{carda} in the context
of the double quantum dot system coupled to a continuum of states.

Due to the presence of non-Hermitian terms, we write the adjoint symmetric and anti-symmetric states
corresponding to the states in Eq.(\ref{Eigen}) as
\bea
\nonumber
\ket{\tilde{\chi}_{\rm s}(t)}  & = &  e^{-\bar{\gamma} t/2} \left(\cos\frac{\Omega}{2} t+i \cos \theta \sin \frac{\Omega}{2} t \right)
 \ket{\bf l} \; \\ \nonumber&&- i  e^{-\bar{\gamma} t/2} \sin \theta \sin\frac{\Omega}{2}  t \ket{\bf m} \\ \nonumber
\ket{\tilde{\chi}_{\rm a}(t)}  & = &  e^{-\bar{\gamma} t/2}  \left(\cos\frac{\Omega}{2} t+i \cos \theta\sin \frac{\Omega}{2} t \right)\ket{\bf l} \;
 \\ \nonumber && +i  e^{-\bar{\gamma} t/2} \sin \theta \sin\frac{\Omega}{2} t \ket{\bf m},
\\
\label{EigenA}
\eea
In the case of strong dissipative processes and at rising temperatures, it is likely that the condition
$2 V_r={\gamma^{\star}}$  will be   satisfied and  $\Omega$=0. This signifies the
appearance of the exceptional point \cite{heiss}  when  both coherent and incoherent tunneling regimes merge,
 and the  two eigenvalues coalesce  to represent  just one eigenfunction. 
We obtain $P_{ll} =  \left(1-  \frac{\gamma^{\star} t}{2}\right)^2 e^{-\bar{\gamma} t}$, 
$ P_{lm} = \left(\frac{\gamma^{\star} t}{2}\right)^2 e^{-\bar{\gamma} t}$ and the population difference $P_{ll}-P_{lm}$=
$\left(1-  \gamma^{\star} t\right) e^{-\bar{\gamma} t}$

\section{\label{bch} Application to the B850 Bchls dimer model}
The LH2  B850 pigment-protein complex which interacts with the reaction center via LH1,
 consists of a cyclic array of polypeptide heterodimers, with each polypeptide pair 
of $\alpha \beta$ subunit, containing
three Bchls. A pair of  Bchls is located near the outer membrane
surface of the complex, while  a single Bchl  is present near the inner membrane  as detailed in Ref.\cite{johan}. 
A similar cyclic structure (B800 ring), but one with a smaller number of  Bchls accounts for the 
800 nm absorption spectrum of the LH2 complex. We utilize  a  simple model which 
consists of two coupled B850 Bchls, each BChl with a two-state structure forming
 the basis of the Q$_y$ one-exciton states.  The   Q$_y$ bandwidth is associated with 
the lowest excited state of the BChl molecule. The  two-state model of the BChl
interacts with its surrounding phonon bath  with properties similar to that of 
a spin-boson system \cite{legg}, and forms an integral part of 
light-harvesting energy transfer mechanisms in the LH2  B850 pigment-protein complex.

We  employ the Hamiltonian of the form given in  Eq. (\ref{exHim}) to represent the B850 Bchls dimer,
with the   intersite  coupling energy $ V$=250 cm$^{-1}$ and consider 
 the two site energies to be equal, that is  $E_l \approx E_m$.
We use the spectral density function associated with the overdamped 
Brownian oscillator \cite{sbook}
\begin{equation}
\label{spectral}
J(\omega) = \frac {2}{\pi} \lambda_b \omega_0 \; \frac{\omega}{\omega^2+\omega_0^2}
\end{equation}
where $\lambda_b$ is the bath reorganization energy  and
 $\omega_0$ is a frequency cutoff for the bath.
We use values of $\lambda_b$= 200 cm$^{-1}$ and 
$\hbar \omega_0$ = 50 cm$^{-1}$ which are acceptable
 estimates in photosynthetic systems \cite{p3,olbri,engel,renrat,johan}.
It is expected that the  bath reorganization energy  $\lambda_b$
has an implicit dependence on temperature, however here we assume
that it remains independent of temperature for the sake of obtaining some
numerical estimates of the dynamics of energy exchange in the B850 Bchls system.

\begin{figure}[htp]
  \begin{center}
\subfigure{\label{aa}\includegraphics[width=4.4cm]{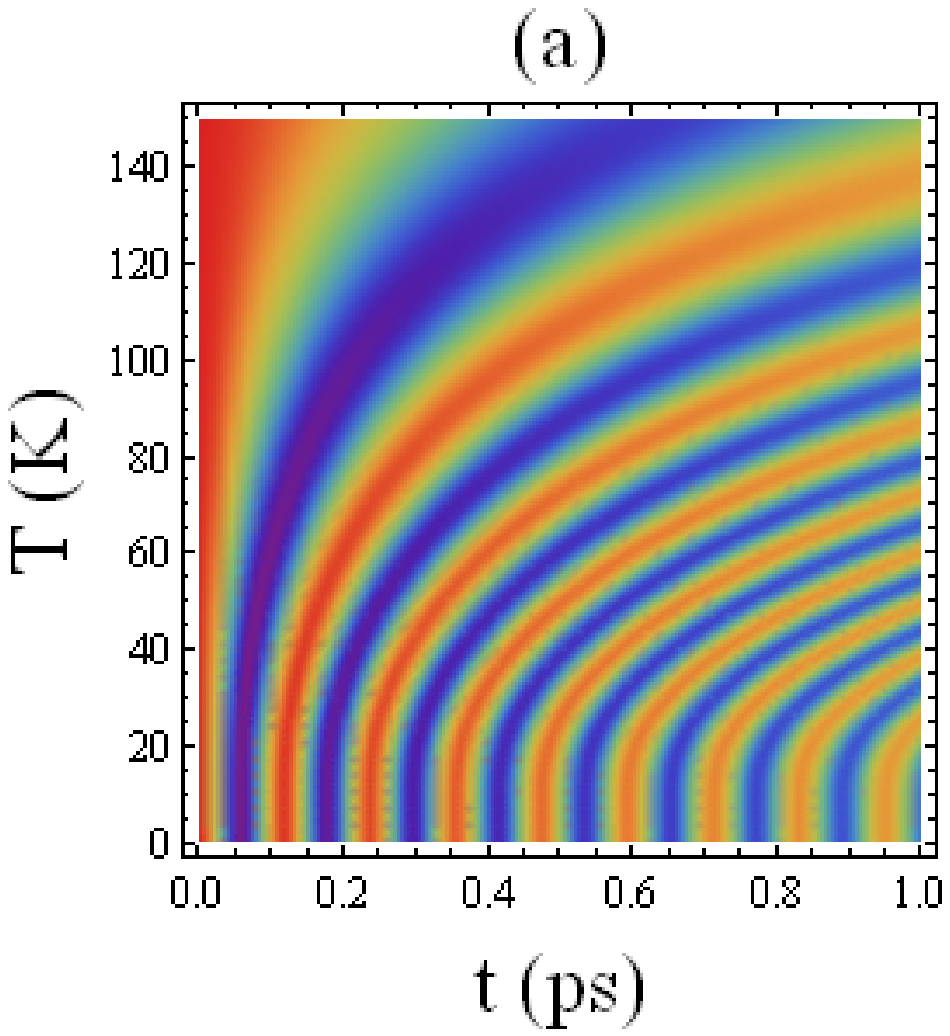}}\vspace{-1.1mm} \hspace{1.1mm}
\subfigure{\label{bb}\includegraphics[width=4.4cm]{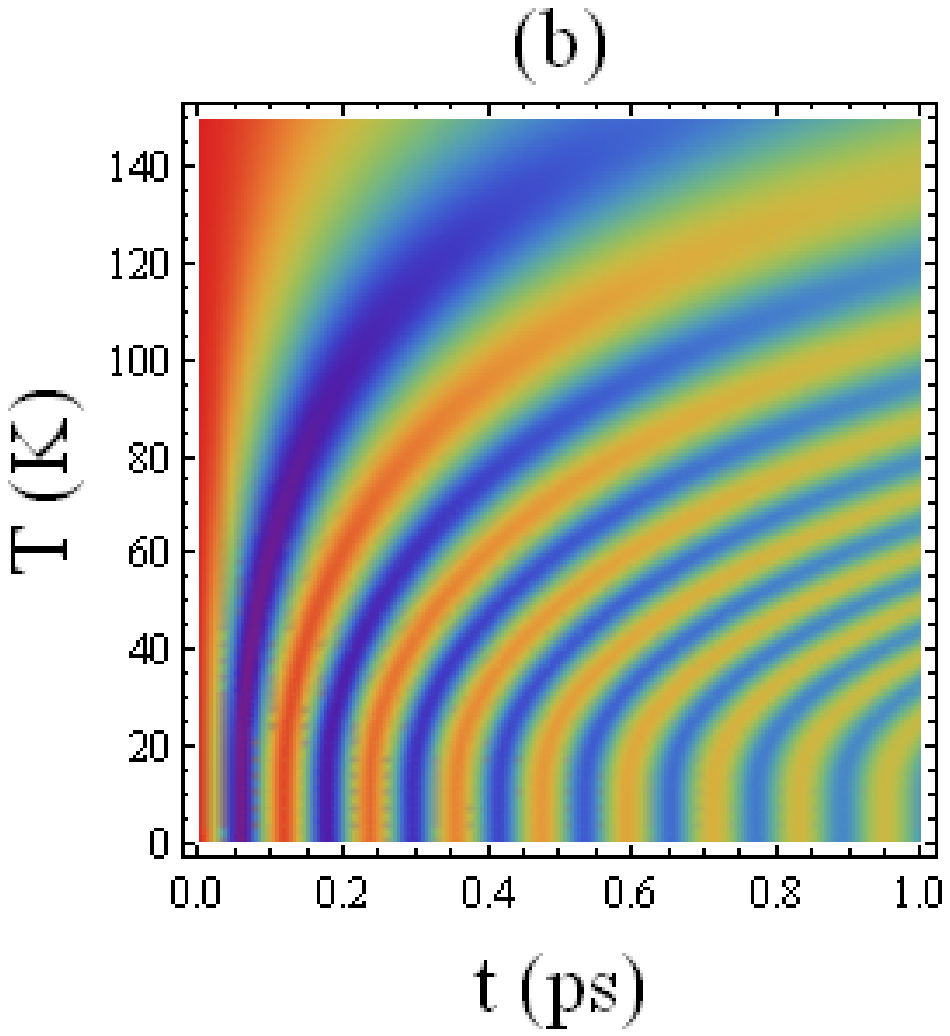}}\vspace{-1.1mm} \hspace{1.1mm}
\subfigure{\label{bb}\includegraphics[width=4.4cm]{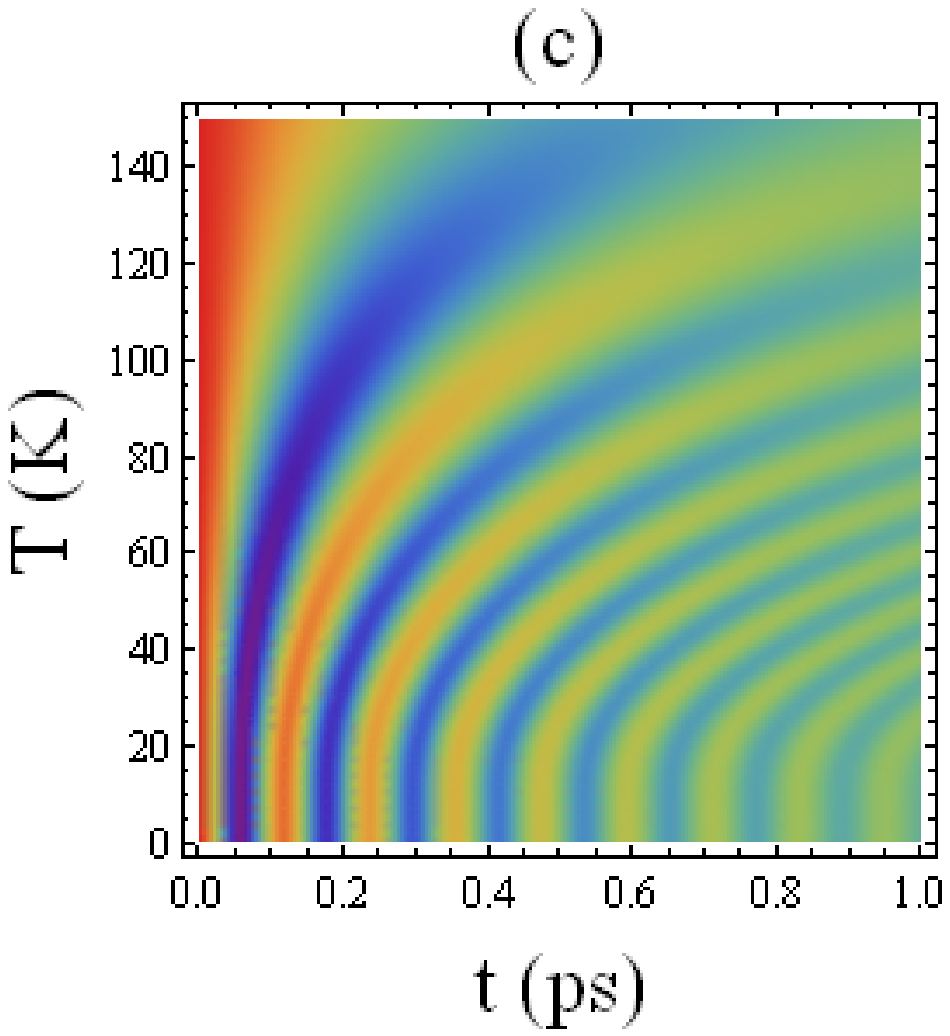}}\vspace{-1.1mm} \hspace{1.1mm}
     \end{center}
  \caption{Fading out of the exciton population difference, $\Delta P$=$P_{ll}-P_{lm}$, with increase in time (ps)  
and temperature (in K) at various dissipation strengths,  $\gamma_m$=$\gamma_l$ = (a) 2.5 cm$^{-1}$  
(b) 5 cm$^{-1}$  (c) 10 cm$^{-1}$. The bath reorganization energy  $\lambda_b$=200 cm$^{-1}$,
$\hbar \omega_0$ = 50 cm$^{-1}$ and bare intersite coupling energy,
$V$=250 cm$^{-1}$ \cite{p3,olbri,engel,johan}.
The colour shading range from red for the maximum population difference of 1 to blue
for the minimum population difference of -1.}
\label{temp1}
\end{figure}

\begin{figure}[htp]
  \begin{center}
\subfigure{\label{aa}\includegraphics[width=4.4cm]{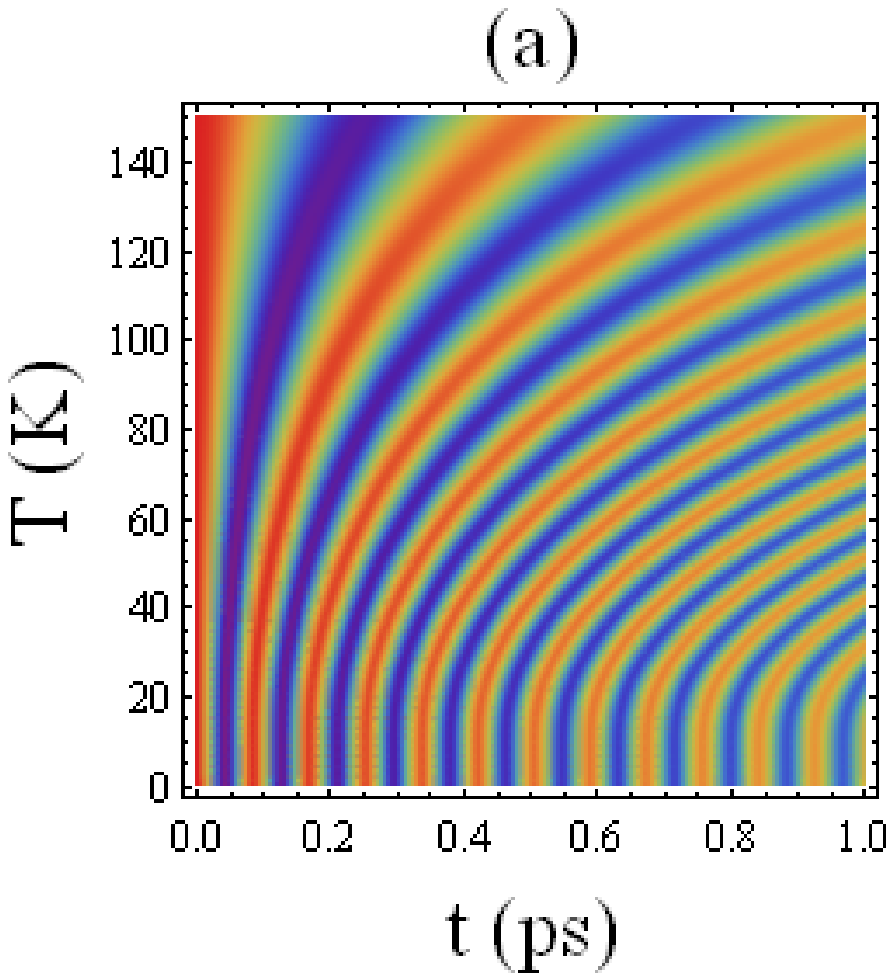}}\vspace{-1.1mm} \hspace{1.1mm}
\subfigure{\label{bb}\includegraphics[width=4.4cm]{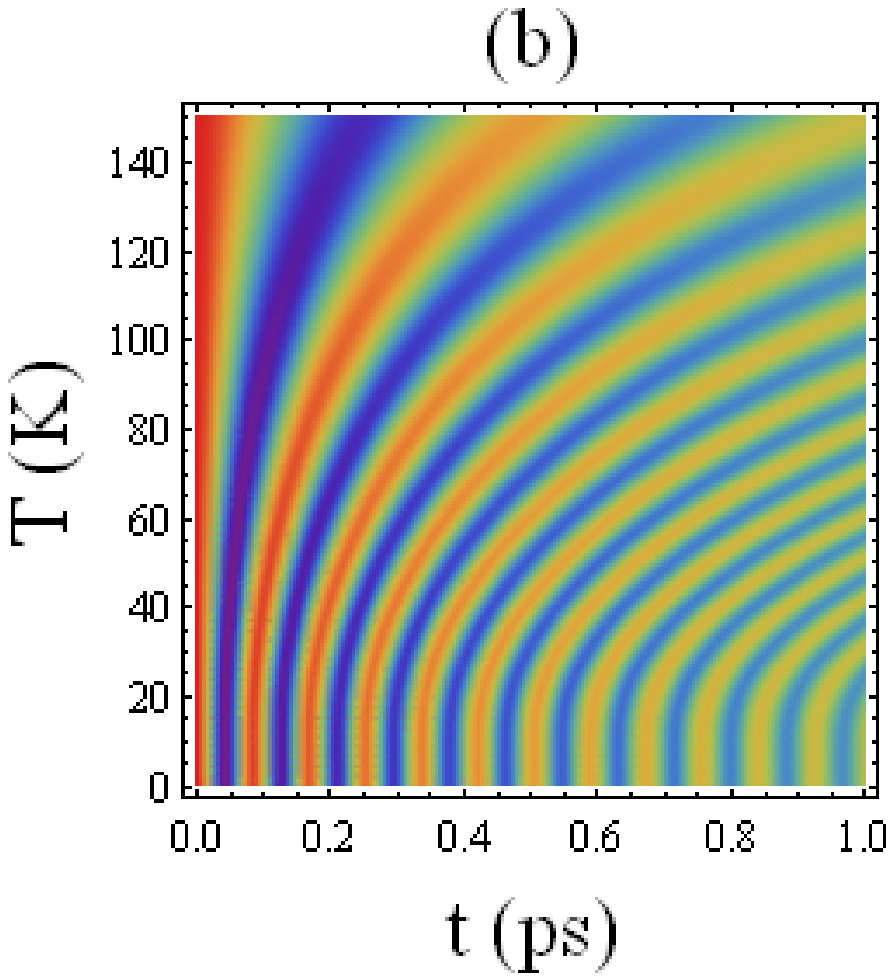}}\vspace{-1.1mm} \hspace{1.1mm}
\subfigure{\label{bb}\includegraphics[width=4.4cm]{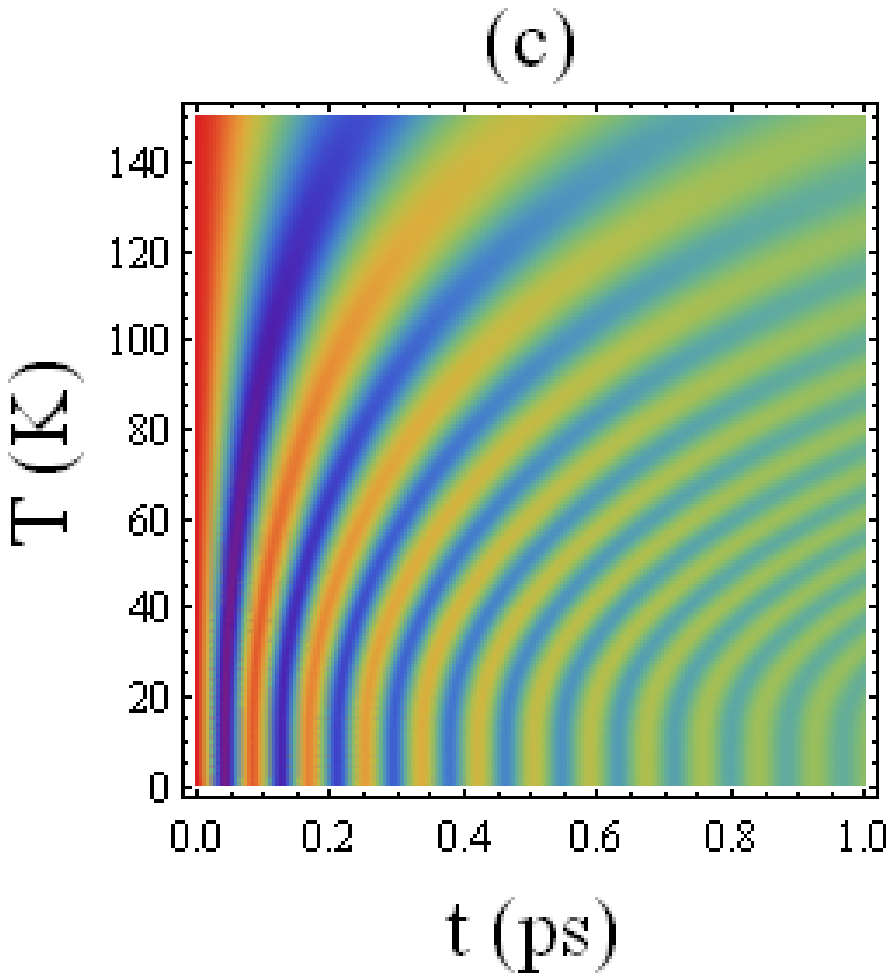}}\vspace{-1.1mm} \hspace{1.1mm}
     \end{center}
  \caption{Population difference, $\Delta P$=$P_{ll}-P_{lm}$ as function of  time (ps)  
and temperature (in K) at various dissipation strengths,  
$\gamma_m$=$\gamma_l$ = (a) 2.5 cm$^{-1}$  (b) 5 cm$^{-1}$  (c) 10 cm$^{-1}$. The bath reorganization energy  is set lower at $\lambda_b$=150 cm$^{-1}$.
All other parameters ($\hbar \omega_0$,$V$) and colour codes are set at the same values
as specified in the caption of  Fig.~\ref{temp1}.}
\label{temp2}
\end{figure}
The  effect of  temperature on the time evolution of the
population difference,$\Delta P$=$P_{ll}-P_{lm}$,  at various dissipation levels
is shown in Fig.~\ref{temp1}a,b,c,
While oscillations are preserved at small dissipation levels up to moderate
temperatures (80 K), there is less exchange of energy between the two sites
at the comparatively larger temperatures (140 K). Moreover the probabilities
of site  occupation $P_{ll}$ and $P_{lm}$ becomes zero due to outflow of
energy to the dissipation sources. We note increased oscillations in $\Delta P$
 as illustrated in Fig.~\ref{temp2}a,b,c when the bath reorganization energy  is reduced to a
lower value of  $\lambda_b$=150 cm$^{-1}$. As mentioned earlier, the results in these figures
are based on the assumption that the reorganization energy is independent of the
temperature, which has been used in  earlier works \cite{renrat,johan}. Currently,
the explicit  dependence of  $\lambda_b$ on the temperature is lacking in the literature.
The effect of including  the temperature dependence of $\lambda_b$ is expected
to alter the quantitative estimates of the timescale of oscillations by a small factor,
however we expect the salient qualitative features to be preserved. It is  to be noted 
that the time estimates obtained using our simple dimer model is in partial agreement
with  experimental results \cite{collini,pani} which show that coherent oscillations persist 
up to 0.3 ps even at physiological temperatures for a similar photosynthetic system.
The oscillations in   population difference $\Delta P$ as illustrated in 
Figs.~\ref{temp1}, ~\ref{temp2} are clearly dependent on the 
strength of exciton coupling to the sources of dissipation and the estimates
in the range (2 to 10) cm$^{-1}$ employed here may be  lower than those 
in  photosynthetic systems. Accordingly,  recombination and trapping effects
which are common sources of dissipation, are expected to have a significant effect on the  photosynthetic activity of
light harvesting systems in general.

\begin{figure}[htp]
  \begin{center}
\subfigure{\label{aa}\includegraphics[width=6.7cm]{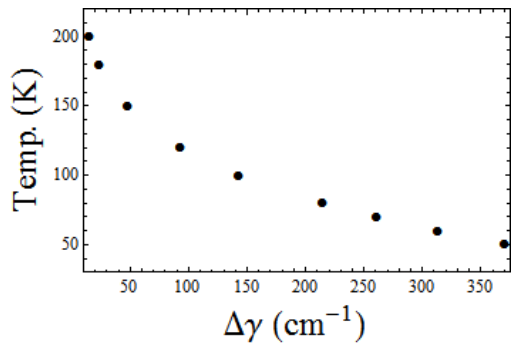}}\vspace{-1.1mm} \hspace{1.1mm}
     \end{center}
  \caption{Critical temperatures (K) at which exceptional points occur,  as a function of the dissipation strength difference,  $\Delta \gamma$=$|\gamma_m$-$\gamma_l|$. 
All other parameters ($\hbar \omega_0$, $\lambda_b$, $V$) are set at the same values
as specified in the caption of Fig.~\ref{temp1}.}
\label{crit}
\end{figure}

Fig.~\ref{crit} illustrates the dependence of the critical temperatures (K)  which occur at exceptional points,  as a function of the dissipation strength difference,  $\Delta \gamma$=$|\gamma_m-\gamma_l|$ based on 
Eq.(\ref{compE}). The results indicate that some degree of coherence can be obtained at
physiologically higher  temperatures ($>$ 70K) as long as the dissipation strength difference,  $\Delta \gamma$
is kept low. It is to be noted that the estimates of the critical temperatures in Fig.~\ref{crit} is based on a simple dimer
model which circumvents comprehensive details that
 incorporate the energy bias  and solvent relaxation rate, various 
 kinetic behaviors of the electron transfer mechanisms and dynamic
bath effects as considered in Ref.\cite{cao}. 
In general, the Redfield model employed here works well as a weak coupling model,  however
 its validity is very much dependent on the partitioning  between  the exciton Hamiltonian and the 
intersite coupling and exciton-phonon terms. In Eq. (\ref{exHim}) we have chosen the exciton states as
the zeroth order term, which may not provide a reliable description of the dynamics in light-
harvesting complexes, under certain environmental conditions \cite{caoIop}.
 It has been noted earlier  that  the secular Redfield equation \cite{hierac}
 leads to an overestimation of the   environment-assisted transfer process which  results in inaccuracies in the prediction
of the coherent-incoherent transition region. One approach  of overcoming this problem  is by   extending
Eq. (\ref{exHim}) to dressed excitons  or composite exciton-phonon states 
with  hybrid exciton-phonon characteristics. This can be done using
suitable canonical transformations of  exciton and phonon operators to a  transformed Hilbert space,
however analytical solutions can only be obtained under strict limits.
While the analytical model  employed here offers advantage in the computational 
speed of prediction of the exception point, we expect some quantitative  changes (less than a order of 
magnitude) in the critical temperatures shown in  Fig.~\ref{crit}  when improved models incorporating  more realistic features are utilized. Nevertheless, the qualitative trends, especially
 changes in  the exceptional point temperatures with dissipation strength differences,
are expected to be unaffected by the assumptions inherent in the Redfield model for the range of $\Delta \gamma$
and  bath reorganization energies considered in this work. In this regard,
 the underlying principles associated with the appearances of exceptional points
can be generalized to studies which include more elaborate system parameters.
Hence there is ample scope to extend the intermediate complexity of the
current work  to  scrutinize the factors which influence 
 the  crossover point from   coherent to incoherent dynamics, and  influence the
 appearance of exceptional points  in more realistic systems with  strong system-bath couplings. 

The connection between the exactly solvable two-state model 
to light-harvesting energy transfer processes should also be viewed in the context
of the  large molecular structure of realistic photosynthetic systems.
Extrinsic factors such as network size and topological connectivity present in large molecular structures
assume vital roles in the optimum energy transfer processes, and  consequently 
determine the  coherence  properties of the entangled exciton in large molecular structures. 
In a recent work \cite{caonet}, the mapping of quantum energy transfer processes to network
kinetics with nonlocal connectivity was examined in three-level systems, linear-chains, and other closed
networks. The  results indicated that environments can be optimized to maximize  energy
transfer efficiency, with prediction of  phase-sensitive interferences
in closed-loop configurations. These factors provide sufficient framework  to examine the appearances
of exceptional points within the structure of large topologically connected structures
and their possible influence on  phase-sensitive interferences \cite{caonet}.

\section{\label{brac} The quantum brachistochrone passage times in photosynthetic systems.}
\begin{figure}[htp]
  \begin{center}
\subfigure{\label{aa}\includegraphics[width=8cm]{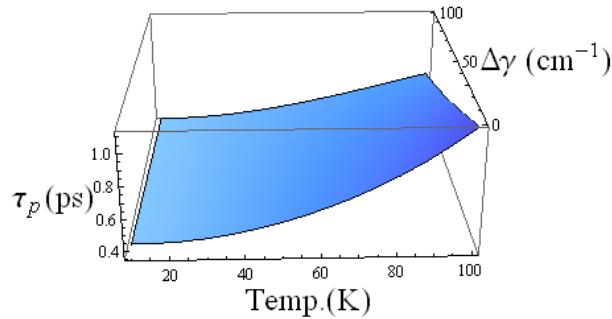}}\vspace{-1.1mm} \hspace{1.1mm}
     \end{center}
  \caption{Passage time from the $\ket{\bf l}$ to  $\ket{\bf m}$  states 
as a function of the dissipation strength difference,  $\Delta \gamma$=$|\gamma_m$-$\gamma_l|$
and temperature. The bath reorganization energy  $\lambda_b$=150 cm$^{-1}$,
$\hbar \omega_0$ = 50 cm$^{-1}$ and bare intersite coupling energy,
$V$=250 cm$^{-1}$}
\label{pass}
\end{figure}

The exact role of exceptional points in photosynthetic systems remains unclear, and it is 
likely that these points are  linked to the  quantum non-Hermitian brachistochrone problem \cite{ali,bender}.
In the general brachistochrone problem, the minimum time taken to transverse the 
path  between two locations of a particle is  determined, this  becomes the
quantum brachistochrone problem when the time of evolution of  the quantum system
between two states is required. It has been shown that the passage time of  evolution of an
initial state  into the final state can be made arbitrarily small for
a time-evolution operator  that is  non-Hermitian but {\it PT}-symmetric \cite{bender}. This result
has recently been generalized  to  non-{\it PT}-symmetric dissipative
systems \cite{fring}. The results in Refs. \cite{bender,fring} appear to suggest that 
the timescales of propagation in non-Hermitian quantum mechanics  are faster than those
of  Hermitian systems. Using the approach detailed in Ref. \cite{fring}, we compute the 
passage time, $\tau_p$ taken for the $\ket{\bf l}$ state to make a transition to the
$\ket{\bf m}$ state as a function of temperature and the dissipation strength difference,  
$\Delta \gamma$=$|\gamma_m$-$\gamma_l|$ (Fig.~\ref{pass}). 

As expected, $\tau_p$ increases with temperature, however it decreases
with the dissipation strength difference,  $\Delta \gamma$.  The passage times (in ps) obtained here are  small compared to typical estimates (1 ns) of the fluorescence  decay times of the light harvesting complex \cite{pala}.
This suggest considerable delocalization, with implications for 
non-trivial quantum correlations amongst various pigment sites. However the explicit 
link between the entanglement properties and the brachistochrone passage times
is not well understood.  There is some consistency of the results in Fig.~\ref{pass}
 with earlier results which show loss  of coherence at
physiologically higher  temperatures (Fig.~\ref{crit}), however the decrease of $\tau_p$
with  the dissipation strength difference,  $\Delta \gamma$=$|\gamma_m$-$\gamma_l|$
may be associated with a decrease in efficiency of the energy transfer mechanism.
Clearly there are open questions with regard to the entanglement properties of quantum systems
which obey non-Hermitian quantum dynamics, these properties are  comparatively well
established in Hermitian systems. For instance there  is lack of consensus on the 
properties of the non-Hermitian eigenstates, as while it is known that real eigenstates
are associated with localized states, eigenvalues of non-Hermitian eigenstates may have
complex delocalization   features which are not well understood \cite{hatano}.

The results obtained here can be compared to earlier works \cite{ok1,ok2,ok3} of coherent exciton dynamics
 which show that the environment not only allow the preservation of some coherent
features but also  assist in the quantum transport  in light-harvesting systems, even at ambient temperatures.
However the role of exceptional points have not been implicated in the earlier works, and in this respect,
the estimates of the quantum brachistochrone passage times here is expected to bring greater depth to existing
studies in this field. Hence the  solvable model examined in this work  suggest features that have not been predicted
in earlier works.  Nevertheless, the role of new 
measures of quantum correlations  which  possess more generalized properties
such as  the quantum discord  \cite{zu,ve1} in the photosynthetic process needs further
investigation, including examination of the intricate links between the exceptional point
and  quantum correlation measures. 

Lastly, we have presented calculations associated with a  dissipative two-level dimer model  which 
incorporates  decay terms due to  coupling to a macroscopic environment. The
results when applied to the B850 Bchls dimer system  can account partly for the observed
high efficiencies of energy transfer in light harvesting systems,
and  highlight the critical link between photosynthetic coherences
and  dissipation strength differences at the two sites of the dimer.
There is greater efficiency in energy transfer 
as long as the distinguishability of the subunits of the dimer is minimized,
an analogous observation was  made for the system  of quantum dot systems 
connected to conducting leads in an earlier work\cite{carda}.
Thus in the  presence of a small  dissipation strength difference,  $\Delta \gamma$
the excitonic coherences appear  insensitive to the environmental changes, provided
the subunits are not distinguished by varying the strength of dissipation.
The most important result of this study is however related to the appearance of the exceptional
point which has not received attention in earlier works on photosynthetic systems.
Despite the frequent mention of the crossover point from   coherent to incoherent dynamics
at increasing temperatures in the literature, the presence of the exceptional points has been overlooked.
The evaluation of the quantum brachistochrone passage times is also unique to this work.
However  the  role of exceptional points need further investigations, as there are many unanswered
questions as to how such points actually related to the high efficiencies measure in
light harvesting systems.

It is to be noted that the model used in this work can be applied to obtain numerical estimates of
quantum coherence in other  photosynthetic systems which possess more realistic
spectral functions associated with the environmental couplings . 
The model used here can also be extended to examine the properties of the
exceptional points and the quantum brachistochrone passage times
in a dimer model where the onsite and coupling energies differ as well.
It is expected that the qualitative features involving the appearances of exceptional points,
and shown within the intermediate complexity of the current work, 
are expected to be retained in more sophisticated models.
 In this regard, the results obtained in the current work provide important guidelines 
for future investigations in more sophisticated
models, and in the detection and control
of quantum effects in artificial photosynthetic systems.

\section{Acknowledgments}

This research was undertaken on the NCI National Facility in Canberra, Australia, which is
supported by the Australian Commonwealth Government.


\begin{thebibliography}{99}

\bibitem{p0}
J. J. Hopfield, Proc. Nat. Acad. Sci. U.S.A.,
{\bf  71}, 3640 (1974).


\bibitem{fors}
Th. Förster, Ann. Phys. {\bf 437}, 55 (1948).


\bibitem{p1}
H. Zuber and R. Cogdell, in {\it Anoxygenic Photosynthetic Bacteria}, edited by
R. Blankenship, M. Madigan, and C. Bauer, pp. 315–348 (Kluwer Academic, Dordrecht,
1995).

\bibitem{p1a}
B. Chance and M. Nishimura, Proc. Nat. Acad. Sci. U.S.A., {\bf 46}, 19 (1960).


\bibitem{p2}
R. S. Knox, in {\it  Primary Processes of Photosynthesis}, edited by J. Barber,
pp. 55–97 (Elsevier, Amsterdam, 1977).

\bibitem{p3}
{\it Light-Harvesting Antennas in Photosynthesis}, edited by B. R. Green and
W. W. Parson (Springer, New York, 2003).

\bibitem{may}
V. May and O. K\"uhn, {\it Charge and Energy Transfer Dynamics in Molecular
Systems}, 2nd ed. (Wiley-VCH, Berlin, 2004).

\bibitem{macro}
B. Happ, J.  Sch\"afer, R.  Menzel, M. D Hager, A.  Winter,
J. Popp, R. Beckert, B. Dietzek and U. S. Schubert,
Macromolecules,  {\bf 44}, 6277, (2011).

\bibitem{lloyd}
M.  Mohseni, P.  Rebentrost, S.  Lloyd and A. Aspuru-Guzik, J. Chem. Phys. {\bf 129}, 174106 (2008).


\bibitem{silbey}
E. N. Zimanyi and R. J. Silbey, J. Chem. Phys.{\bf 133}, 144107 (2010).


\bibitem{ghosh}
P. K. Ghosh, A. Y.  Smirnov and F. Nori, J. Chem. Phys.{\bf 134}, 244103 (2011).



\bibitem{olbri}
C. Olbrich, T. L. C. Jansen, J.  Liebers, M.  Aghtar, J.  Str\"umpfer. K. Schulten, 
J.  Knoester, and U.  Kleinekath\"ofer, J. Phys. Chem. B,  {\bf  115}, 8609 (2011).

\bibitem{pala1}
H.  van Amerongen and R. van Grondelle,
J. Phys. Chem. B {\bf  105}, 604 (2001).

\bibitem{pala2}
P. Horton, and A. V. Ruban, J. Exp. Bot. {\bf 56}, 365 (2005).

\bibitem{pala}
M.  A. Palacios, F. L. de Weerd, J.  A. Ihalainen,  R. van Grondelle, and
H.  van Amerongen, J. Phys. Chem. B {\bf  106},  5782 (2002).


\bibitem{flem}
T. Brixner, J. Stenger, H. M. Vaswani, M.  Cho,  R. E.  Blankenship, and 
G. R.  Fleming,  Nature,   {\bf  434}, 625 (2005).


\bibitem{fenn}
R. E. Fenna and B. W. Matthews, Nature, {\bf 258}, 573 (1975).


\bibitem{engel}
G. S. Engel, T. R. Calhoun, E. L. Read, T.K. Ahn, T. Mancal, Y. C. Cheng,
R. E. Blankenship, and G. R. Fleming, Nature, {\bf  446}, 782 (2007).


\bibitem{renrat}
N. Renaud, M. A. Ratner and V. Mujica, 
J. Chem. Phys. {\bf 135},  075102 (2011).

\bibitem{reng}
J. Adolphs and T. Renger,
Biophys. J. {\bf 91}, 2778 (2006).

\bibitem{brug}
B. Bruggemann, P. Kjellberg, and T. Pullerits, Chem. Phys. Lett. {\bf  444}, 192
(2007).


\bibitem{ener1a}
R. Siebert, A. Winter, B. Dietzek, U. Schubert, J. Popp, 
J. Macromol. Rapid Commun., {\bf 31}, 883 (2010).

\bibitem{ener1b}
J. Schaefer, R. Menzel, D. Weiss, B. Dietzek, R. Beckert, J. Popp,
 Journal of Luminescence, {\bf  131}, 1149 (2011).

\bibitem{ener1c}
H. D\"urr and S. Bossmann, Acc. Chem. Res., {\bf 34}, 905 (2001).

\bibitem{ener1}
W. D. Larkum, Curr. Opin. Biotechnol. {\bf 21}, 271 (2010).

\bibitem{ener2}
Y. Terazono, G. Kodis, P. A. Liddell, V. Garg, T. A. Moore, A. L. Moore,
and D. Gust, J. Phys. Chem. B, {\bf  113}, 7147 (2009).

\bibitem{ener3}
R. M. Clegg, M. Sener and  Govindjee, 
Proc. SPIE {\bf 7561}, 75610C (2010)

\bibitem{ener4}
D. Gust, T. A. Moore, and A. L. Moore,
Acc. Chem. Res., {\bf  34}, 40 (2001).

\bibitem{collini}
E. Collini, C. Y. Wong, K. E. Wilk, P. M.G. Curmi, P. Brumer, 
and G. D. Scholes, Nature, {\bf 463}, 644 (2010).


\bibitem{flemB}
M. Cho,  H. M. Vaswani, T. Brixner,  J. Stenger, and
G. R.  Fleming,  J. Phys. Chem. B {\bf 109}, 10542 (2005).


\bibitem{pani}
G. Panitchayangkoon, D. Hayes, K. A.  Fransted, J. R.  Caram, E.  Harel, J. Wen, 
R.  Blankenship, G. S.  Engel,  Proc. Nat. Acad. Sci. {\bf 107}, 12766 (2010).



\bibitem{segale}
D. Segale and V. A. Apkarian,
J. Chem. Phys. {\bf 135},  024203 (2011).



\bibitem{galve}
F. Galve, L. A. Pachon and D. Zueco,
Phys. Rev. Lett. {\bf 105}, 180501 (2010).


\bibitem{fas}
F. Fassioli and A. Olaya-Castro,  New J. Phys. {\bf 12}, 085006 (2010).


\bibitem{car}
F. Caruso,  A. W. Chin, A. Datta, S. F.  Huelga and M. B. Plenio, Phys. Rev. A {\bf  81}, 062346 (2010).

\bibitem{sar}
M. Sarovar, A.  Ishizaki,  G. R. Fleming and K. B. Whaley, K. B., Nature Physics {\bf 6}, 462 (2010).

\bibitem{scholak}
T. Scholak, F.  de Melo, T.  Wellens, F.  Mintert, A. Buchleitner, Phys. Rev. E {\bf 83}, 021912 (2011).


\bibitem{nazir}
 A. Nazir,  Phys. Rev. Lett.  {\bf 103}, 146404 (2009).

\bibitem{fass}
F. Fassioli, A. Nazir, A. J.  Olaya-Castro, Phys. Chem. Lett. {\bf 1}, 2139 (2010).

\bibitem{nal}
P. Nalbach, J. Eckel, M. Thorwart,  New J. Phys. {\bf  12}, 065043 (2010).


\bibitem{red}
A.G. Redfield, Advances in Magnetic Resonance (Academic Press Inc., New York, 1965), {\bf  1}, pp. 1–32.


\bibitem{weiss} 
 U. Weiss, {\it Quantum Dissipative Systems}, 
  (World Scientific, Singapore, 1993).   


\bibitem{breu}
H. P. Breuer and F. Petruccione, {\it The Theory of Open Quantum
Systems } (Oxford University Press, New York, 2002).

\bibitem{pot}
E. O. Potma and D. A. Wiersma, J. Chem. Phys., {\bf 108}, 4894 (1998).


\bibitem{hierac}
A. Ishizaki and G. R. Fleming, J. Chem. Phys. {\bf 130}, 234111 (2009).

\bibitem{caoIop}
J. Wu, F. Liu, Y.  Shen, J.  Cao and R. J. Silbey, New J. Phys. {\bf 12}, 105012 (2010).


\bibitem{shaun}
D. Abramavicius and S. Mukamel, J. Chem. Phys.  {\bf 133}, 064510 ( 2010).


\bibitem{johan}
J. Str\"umpfer and K. Schulten, J. Chem. Phys. {\bf 134}, 095102 (2011).


\bibitem{econ}
E. N. Economou, {\it Green's Functions in Quantum Physics}
(Springer-Verlag, Berlin, 1979). 
   


\bibitem{mal}
J. Malinsky and Y. Magarshak, J. Phys. Chem., {\bf 96} ,2849 (1992).

   
\bibitem{suna}
A. Suna, Phys. Rev. {\bf 135}, A111 (1964).


\bibitem{Dav}
A. S. Davydov, {\it Theory of Molecular Excitons} (Plenum, New York, 1971).


\bibitem{thila}
A. Thilagam,  J. Phys. A: Math. Theor. {\bf 43} 155301 (2011).


\bibitem{tell}
E.  Teller, J. Phys. Chem. {\bf  41}, 109 (1937).


\bibitem{heiss}
W. D. Heiss and A. L. Sannino, Phys. Rev. A {\bf 43},  4159 (1991).


\bibitem{thilaberry}
A. Thilagam,  J. Phys. A: Math. Theor. {\bf 43} 354004 (2010).


\bibitem{microexpt}
C. Dembowski, B. Dietz, H. D. Gr\"{a}f, H. L.  Harney, A.  Heine, 
W. D.  Heiss and A. Richter 2001 Phys. Rev. Lett.
{\bf 90},  034101 (2003).


\bibitem{b1} B. Dietz, T. Friedrich, J. Metz, M. Miski-Oglu, A.
Richter, F. Sch\"{a}fer, and C. A. Stafford, Phys. Rev. E {\bf 75}, 027201
(2007).

\bibitem{b2} C. Dembowski,B. Dietz, H. D. Gr\"{a}f, H. L. Harney,
A. Heine, W. D. Heiss, and A. Richter, Phys. Rev. E {\bf 69}, 056216
(2004).


\bibitem{cav} G. Khitrova, H. M. Gibbs, M. Kira, S. W. Koch, 
and A. Scherer, Nat. Phys.{\bf 2}, 81 (2006).


\bibitem{lefe}
R. Lefebvre, O. Atabek, M. Sindelka, and N. Moiseyev,
Phys. Rev. Lett. {\bf 103}, 123003 (2009).


\bibitem{lefe2}
R. Uzdin and R. Lefebvre, J. Phys. B: At. Mol. Opt. Phys.  {\bf 43}, 235004 (2010).

\bibitem{carta}
H. Cartarius, J. Main, and G. Wunner, Phys. Rev. Lett. {\bf 99},
173003 (2007).

\bibitem{legg} A. J. Leggett, S. Chakravarty. A. T. Dorsey, M. P. A. Fisher,
A. Garg, and W. Zwerger, Rev. Mod. Phys. {\bf 59}, 1 (1987).

\bibitem{statt}
 C. A. Stafford  and B. R.  Barrett, 
{\it Phys. Rev. C} {\bf 60}  051305, (1999).



\bibitem{mah}
G. D. Mahan, {\it Many Particle Physics}, (Plenum, New York, 2000).



\bibitem{Thilaprb}
A. Thilagam, Phys. Rev. A {\bf 81}, 032309 (2010).


\bibitem{carda}
D. M. Cardamone, C. A. Stafford, and B. R. Barrett,
Phys. stat. sol. (b), {\bf  230}, 419  (2002).


\bibitem{sbook}
S. Mukamel, {\it Principles of Nonlinear Optics and Spectroscopy} (Oxford University Press,
New York, 1995).

\bibitem{cao}

Y. J. Jung and J. Cao,  J. Chem. Phys. {117}, 3822 (2007).

\bibitem{caonet}
J. Cao and R. J. Silbey, J. Phys. Chem. A, {\bf  113}, 13825 (2009).


\bibitem{ali}
 A. Mostafazadeh,  Phys. Rev. Lett.  {\bf 99} 130502 (2007).

\bibitem{bender}
 C. M. Bender, D. C.  Brody, H. F. Jones and  B. K.  Meister,
 Phys. Rev. Lett. {\bf 98}, 040403 (2007).

\bibitem{fring}
P. E. Assis and A. Fring, J. Phys. A: Math. Theor. {\bf 41}, 244002 (2008).

\bibitem{hatano}
 N. Hatano and D. R. Nelson,  Phys. Rev. B {\bf 58}, 8384 (1998).

\bibitem{ok1}
F. Caruso,  A. W. Chin, A. Datta, S. F.  Huelga and M. B. Plenio, J. Chem. Phys.  {\bf  131}, 105106 (2009).

\bibitem{ok2}
 P.  Rebentrost, M.  Mohseni, I. Kassal, S.  Lloyd and A. Aspuru-Guzik, New. J.  Phys. {\bf 11}, 033003  (2009).

\bibitem{ok3}
O. M\"ulken and T. Schmid, Phys. Rev. E {\bf 82}, 042104 (2010).

\bibitem{zu} H. Ollivier and W. H. Zurek, Phys. Rev. Lett. {\bf 88}, 017901 (2001).

\bibitem{ve1} L. Henderson and V. Vedral, J. Phys. A {\bf 34}, 6899 (2001).


  
\end{thebibliography}
\end{document}